# 17

# FRT Regulation in China

*Jyh-An Lee and Peng Zhou*

## 17.1 INTRODUCTION

Facial recognition technology (FRT) applications enjoy a staggering level of penetration in China. Valuing the technology's function in facilitating social control and public security, the Chinese government has not only implemented it widely,[1] but also used it to build a national surveillance architecture together with other mechanisms, such as the social credit system.[2] When providing telecommunications, banking, and transportation and other services, an increasing number of state-owned enterprises record citizens' facial data for their FRT systems.[3] FRT-empowered applications are also commonly adopted in the private sector,[4] for functions such as online payment, residential security, and hospital checking in.[5] The rapid development and wide adoption of FRT has made China a global leader in this field. In a recent round of the 1:N section of the US National Institute of Standard and Technology's (NIST's) Face Recognition Vendor Test, where algorithm providers compete for accuracy, the Hong Kong-based industry giant SenseTime came out on top, together with another China-based service provider.[6] SenseTime, as

---

[1] See, e.g., Seungha Lee, 'Coming into focus: China's facial recognition regulations' (4 May 2020), Center for Strategic & International Studies, www.csis.org/blogs/trustee-china-hand/coming-focus-chinas-facial-recognition-regulations.
[2] Qingxiu Bu, 'The global governance on automated facial recognition (AFR): Ethical and legal opportunities and privacy challenges' (2021) 2 *Int. Cybersecurity L Rev.* 113–145, at 130.
[3] See Yan Luo and Rui Guo, 'Facial recognition in China: Current status, comparative approach and the road ahead' (2021) 25 *U. Pa. J.L. & Soc. Change* 153–179, at 160–162.
[4] See Masha Borak, 'Facial recognition is used in China for everything from refuse collection to toilet roll dispensers and its citizens are growing increasingly alarmed, survey shows' (27 January 2021), *South China Morning Post*, www.scmp.com/tech/innovation/article/3119281/facial-recognition-used-china-everything-refuse-collection-toilet.
[5] Tristan G. Brown, Alexander Statman, and Celine Sui, 'Public debate on facial recognition technologies in China' (Summer 2021), *MIT Case Studies in Social and Ethical Responsibilities of Computing*, https://doi.org/10.21428/2c646de5.37712c5c.
[6] See Chris Burt, 'Top performing developers steady in updated NIST facial recognition 1:N test results' (4 May 2022), BiometricUpdate.com, www.biometricupdate.com/202205/top-performing-developers-steady-in-updated-nist-facial-recognition-1n-test-results.

Asia's largest artificial intelligence (AI) software company, has 22 per cent share of China's computer-vision market.[7] Moreover, surveillance camera makers, such as Hangzhou Hikvision Digital Technology, Zhejiang Dahua Technology, and Megvii Technology, are also leaders in the industry and provide essential equipment for China's pervasive implementation of FRT.[8]

FRT has triggered serious privacy concerns in many countries, and China is of no exception. Although some commentators indicate that Chinese culture is more tolerant towards privacy violations than that of Western countries and many Chinese favour FRT because of increased security or convenience,[9] there have been extensive debates concerning the justification and proper scope of FRT adoption in the country. China has been working on developing a regulatory framework for FRT since 2020. Although this framework aimed to substantially enhance personal data protection, there have been increasing risks and challenges to protect citizens' data in the FRT environment.

This chapter first introduces China's legal framework regulating FRT and analyses the underlying problems. Although current laws and regulations have restricted the deployment of FRT under some circumstances, these restrictions may function poorly when the technology is installed by the government or when it is deployed for the purpose of protecting public security. We use two cases to illustrate this asymmetric regulatory model, which can be traced to systematic preferences that existed prior to recent legislative efforts advancing personal data protection. Based on these case studies and evaluation of relevant regulations, this chapter explains why China has developed this distinctive asymmetric regulatory model towards FRT specifically and personal data generally.

## 17.2 REGULATING FRT IN A FISHBOWL SOCIETY

Given China's over-arching national security drive built on a strong state-centric approach to data governance, its turn to strengthen personal information protection can be somewhat of a puzzle.[10] Heavy investment in FRT and the extensive use by the Chinese government in security applications often portray an invasively transparent 'fishbowl society' straight from Orwellian nightmares.[11] Although the move to more robust protection of personal information appears to conflict with this perception, China has provided an interesting example regarding how authoritarian states balance their digital surveillance and the protection of individuals' personal data. The case of FRT regulations and their enforcement is a particular case to illustrate the challenges of maintaining this balance in China.

---

[7] See Daniel Ren, 'AI, machine learning tech promises US$6000 billion annually for China economy as it pervades industries, says McKinsey' (25 July 2022), *South China Morning Post*, www.scmp.com/business/banking-finance/article/3186409/ai-machine-learning-tech-promises-us600-billion-annually.
[8] Ibid.
[9] Ibid.
[10] Ngoc Son Bui and Jyh-An Lee, 'Comparative cybersecurity law in socialist Asia' (2022) 55 *Vand. J. Transnat'l L.* 631–680, at 660–662.
[11] See Jonathan Turley, 'Anonymity, obscurity, and technology: Reconsidering privacy in the age of biometrics' (2020) 100 *B.U. L. Rev.* 2179–2261, at 2185–2186.

### 17.2.1 *National Laws and Judicial Interpretations*

As early as 2012, the Standing Committee of the Eleventh People's Congress, which is China's top legislative authority, declared its determination to protect digital privacy and planned to legislate data protection principles, such as specific limitations to the collection of personal information and other necessary precautions to safeguard privacy.[12] The 2020 PRC Civil Code (the Civil Code) marked a major shift to the regulatory landscape for the protection of personal information, including biometric data.[13] Prior to the Civil Code, China had no laws regulating FRT. Piecemeal regulations on personal data protection were scattered mostly under laws addressing cyber-crime and cyber-security breaches.[14] The Civil Code dedicates a new chapter to Chinese privacy laws and views personal information as a basic civil right (with the first clause declaring such right in the General Provisions of the Civil Law that came in 2017, as an interim step towards the Civil Code).[15] Article 1035 of the Civil Code establishes general data protection principles, such as purpose and scope limitations as well as the requirement for informed consent by data subjects in processing personal information.[16]

Following the Civil Code, the Supreme People's Court issued the Judicial Interpretation on the Regulation of FRT (the Judicial Interpretation) in 2021.[17] The Judicial Interpretation confirms that facial data falls within the scope of biometrically identifiable information, a type of personal information, prescribed by

---

[12]　Quanguorenmin Daibiaodahui Changwuweiyuanhui Guanyu Jiaqiang Wangluoxinxibaohu de Jueding (《全国人民代表大会常务委员会关于加强网络信息保护的决定》) [Decision of the Standing Committee of the National People's Congress on Strenghening Information Protection on Networks] (2012). Issued by the Standing Committee of the National People's Congress, on 28 December.

[13]　Zhonghua Renmin Gongheguo Minfadian (《中华人民共和国民法典》) [Civil Code of the People's Republic of China (Civil Code)] (2020). Promulgated by the Standing Committee of the National People's Congress on 28 May, effective on 1 January 2021 (hereafter Civil Code), Art. 1034.

[14]　See, e.g., Zhonghua Renmin Gongheguo Wangluo Anquan Fa (《中华人民共和国网络安全法》) [Cybersecurity Law of the People's Republic of China] (2016). Promulgated by the Standing Committee of the National People's Congress on 7 November, effective on 1 June 2017), Art. 41.

[15]　Civil Code, Chapter 6; Zhonghua Renmin Gongheguo Minfa Zongze (《中华人民共和国民法总则》) [General Provisions of the Civil Law of the People's Republic of China] (2017). Promulgated by the Standing Committee of the National People's Congress on 15 March, effective on 1 October 2017, Art. 111.

[16]　Civil Code, Art. 1035.

[17]　Zuigao Renmin Fayuan Guanyu Shenli Shiyong Renlian Shibie Jishu Chuli Geren Xinxi Xiangguan Minshi Anjian Shiyong Falu Ruogan Wenti De Guiding (《最高人民法院关于审理使用人脸识别技术处理个人信息相关民事案件适用法律若干问题的规定》) [Provisions of the Supreme People's Court on Several Issues concerning the Application of Law in the Trial of Civil Cases Relating to Processing of Personal Information by Using the Facial Recognition Technology] (2021). Promulgated by the Judicial Committee of the Supreme People's Court on 8 June, effective on 1 August 2021 (hereafter FRT Judicial Interpretation).

Article 1034 of the Civil Code.[18] Article 2 of the Judicial Interpretation specifically forbids the use of the technology by 'information processors' in public spaces such as hotels, shopping malls, and airports, unless otherwise authorised by authorities.[19] As a reflection of widespread use of facial scanning for identity verification and authentication purposes on residential and commercial properties, Article 10 forbids using FRT without individual consent.[20] The Judicial Interpretation also strengthened remedies for data subjects, including monetary damages and injunctive relief.[21] According to Article 5 of the Judicial Interpretation, liability can be exempted under some circumstances, such as on public security grounds.[22]

Shortly afterwards, the Standing Committee of the National People's Congress passed the PRC Personal Information Protection Law (the PIPL), with a focus on the obligations and liabilities of 'personal information processors' (PIPs).[23] Article 33 stipulates that rules under the PIPL apply to state agencies as well.[24] Moreover, the PIPL views biometric data as a type of 'sensitive personal information',[25] and the processing of such information is subject to a higher standard of protection. PIPs have to obtain independent 'opt-in' consent from data subjects to process such information and inform the latter of the necessity of processing measures as well as the impact on their rights.[26] For individuals under the age of fourteen, such consent must be obtained from parents or statutory agents.[27] Notably, the law allows image collection and personal identification equipment in public places for the purpose of safeguarding public security.[28] Thus, this rule provided a legal basis for security cameras widely deployed by the government.

Several local governments' metropolises have since introduced regulations at provincial and municipal levels to target more narrowly defined scenarios of FRT applications, such as for identity verifications on residential properties.[29] The Municipal

---

[18] Ibid., Art. 1.
[19] Ibid., Art. 2.
[20] Ibid., Art.10.
[21] Ibid., Art. 8 and Art.9.
[22] Ibid., Art. 5.
[23] Zhonghua Renmin Gongheguo Geren Xinxi Baohufa (《中华人民共和国个人信息保护法》) [Personal Information Protection Law of the People's Republic of China (PIPL)]. Promulgated by the Standing Committee of the National People's Congress on 20 Aug 2021, effective on 1 November 2021 (hereafter PIPL).
[24] Ibid., Art. 33.
[25] Ibid., Art. 28.
[26] Ibid., Art. 29.
[27] Ibid., Art. 31.
[28] Ibid., Art. 26.
[29] See, e.g., Hangzhoushi Wuye Guanli Tiaoli (《杭州市物业管理条例》) [Hangzhou Realty Management Regulation] (Hangzhou, China) (2021). Promulgated by the Standing Committee of People's Congress in Hangzhou on 9 August, effective on 1 March 2022, Art. 50; Shanghai Shi Shuju Tiaoli (《上海市数据条例》) [Shanghai Data Regulation] (2021). Promulgated by the Standing Committee of People's Congress in Shanghai on 25 November, effective on 1 January 2022, Art. 23; Shenzhen Jingji Tequ Shuju Tiaoli (《深圳经济特区数据条例》) [Data Regulations of Shenzhen

Government of Hangzhou, for example, amended its Regulation on Realty Management in 2020, limiting the compulsory collection and verification of biometric data such as facial information on residential and commercial properties.[30]

### 17.2.2 Problems Underlying the Current Regulatory Framework

Although China has adopted many internationally recognised data protection principles in its domestic laws,[31] its laws, regulations, and practices regarding FRT and their impact on personal data protection are still controversial. While the consent of data subject is required for another party's data collection, processing, and use, all these procedures can be omitted in the name of public security. A major challenge for personal data protection, in the context of deploying FRT for security purposes, is that the concept of public security does not seem to have any limit and can be interpreted quite expansively.

Taking the hospitality industry, for example, although the Judicial Interpretation specifically forbids the deployment of FRT in places such as hotels, it allows 'laws and regulations' to override this rule for security reasons.[32] To enforce the real-name registration rules,[33] quite a few local governments have mandated hotels to verify the identity of their guests by deploying FRT systems connected to the police database and scanning their faces at check-ins.[34] Although it is not clear whether the hotels have the legal right to process the facial data of their guests, local governments might take advantage of the vague language of the PIPL and infringe on personal data by interpreting the law in a less protective way. Article 13 of the PIPL allows data processing without the data subject's consent for the purpose of 'fulfilling legal responsibility or obligation'.[35] Local governments can easily argue that requiring

---

    Special Economic Zone] (2021). Promulgated by the Standing Committee of People's Congress in Shanghai on 29 June, effective on 1 January 2022, Art. 19.

[30] Ibid.

[31] See James Y. Wang, 'The best data plan is to have a game plan: Obstacles and solutions to reaching international data privacy agreements' (2022) 28 *Mich. Tech. L. Rev.* 385–419, at 401–444.

[32] See FRT Judicial Interpretation, Art. 1 and Art. 5.

[33] See Jyh-An Lee and Ching-Yi Liu, 'Real-name registration rules and the fading digital anonymity in China' (2016) 25 *Wash. Int'l L.J.* 1–34, at 11–15.

[34] In Hunan Province, for example, according to provincial-level real-name registration measures, hotels are required to deploy police systems (the Lüguanye Zhian Guanli Xinxi Xitong, or Public Security Administration Information System) at check-ins to collect facial data. Failing to comply to these measures would deny guests from staying at hotels. In Yushu City of the Qinghai Province, local police started to upgrade the system with FRT-empowered capabilities in 2019. See Hunan Sheng Luguanye Luke Zhusu Shiming Dengji Guanli Guiding (《湖南省旅馆业旅游住宿实名登记管理规定》) [Provisions on the Administration of Real-Name Registration for the Hospitality Industry in Hunan Province] (2021) Promulgated by the Provincial Public Security Department of Hunan Province on 1 December, effective on 1 January 2022, Art.4; The Paper Government Affairs, Lihaile! Yushushi Lüguan Ruzhu Jiang Kaiqi Shualian Shidai (《厉害了！玉树市旅馆入住将开启"刷脸"时代》) [Amazing! Yushu Hotels Now Use Facial Recognition to Check in Guests], *The Paper* (20 November 2019) www.thepaper.cn/newsDetail_forward_5017320.

[35] See PIPL, Art. 13.

hotels to implement FRT is to 'fulfil its legal responsibility or obligation' regarding real-name registration or sector-specific safety policies. This typical example demonstrates that many of the personal data protection mechanisms regarding FRT provided in the laws and judicial interpretation could in reality function less effectively.

Another problem is the asymmetric regulation of FRT in the public and private sectors. While government agencies ordinarily have more chances to be exempted from personal data liabilities because of public security reasons, their liability for data breach is also lighter than that of private parties. While a private party's data misuse would result in both civil and administrative liabilities,[36] Article 68 of the PIPL indicates that violation of personal data rights by the government only leads to administrative liabilities, which would rely on self-correction measures conducted by state agencies.[37] Under this asymmetric framework, it is not surprising that administrative agencies may weigh their own convenience purpose more than personal data protection and thus use FRT in an unbalanced way. The technology has also been deployed to police individuals, including for minor misbehaviour such as jaywalking or wearing pyjamas in public places.[38] It is even reported that the government has used FRT on toilet paper dispensers installed in public toilets to fight off paper thieves.[39] During the COVID-19 pandemic, FRT was deployed comprehensively to verify identities and to monitor and control virus outbreaks on a regular basis.[40]

## 17.3 CASE STUDIES

In recent years, several FRT-related incidents have caught wide public attention and led to lively debates on the potential harm brought by this technology to society.[41] The most noticeable two cases were both raised by law professors challenging the justification of FRT use in citizens' daily lives. Their outcomes, however, differed significantly. While one professor successfully convinced the court that enterprises could not unilaterally impose FRT on its consumers, the other failed to stop its pervasive use in Beijing metro stations.

---

[36] See FRT Judicial Interpretation, Art. 8; PIPL, Art. 66 and Art. 69.
[37] See PRC PIPL, Art. 68. A recent case might illustrate this point. In April 2022, a member of the Big Data Authority in Henan Province was identified in a scandal linked to illicit tempering of personal information from the 'health code' mobile application to wilfully prevent people from retrieving their money from banks that are involved in financial scams. After a public outcry, people deemed directly responsible, including the person from the Big Data Authority, were given administrative and intra-party sanctions, which cited the authority of both the PRC Law on Administrative Discipline for Public Officials (2020) and the party's disciplinary regulations. See, e.g., Phoebe Zhang, 'China officials who abused health codes to stop bank protests punished' (23 June 2022), *South China Morning Post*, www.scmp.com/news/china/politics/article/3182742/china-officials-who-abused-health-codes-stop-bank-protests.
[38] See, e.g., John Wagner Givens and Debra Lam, 'Smarter cities or Bigger Brother? How the race for smart cities could determine the future of China, democracy, and privacy' (2020) 47 *Fordham Urb. L.J.* 829–882, at 865.
[39] Ibid., 865–866.
[40] See, e.g., Jacques deLisle and Shen Kui, 'China's response to Covid-19' (2021) 73 *Admin. L. Rev.* 19–51, 47–48.
[41] Brown, Statman, and Sui, 'Public debate on facial recognition technologies'.

### 17.3.1 *The Hangzhou Safari Park*

China had its first lawsuit concerning the commercial use of FRT in 2019.[42] Bing Guo, a law professor specialising in data protection law, sued Hangzhou Safari Park (HSP) for illegally imposing FRT-based access control after he purchased the annual pass.[43] The Fuyang District People's Court in Hangzhou ruled that HSP breached its contract with Guo by unilaterally changing its entrance policy.[44] However, the court failed to find any data protection violation because the plaintiff agreed to take a photo when he purchased the pass.[45]

In the second instance, the Hangzhou Intermediate People's Court's viewpoint was more favourable to the plaintiff on HSP's use of his facial data. The court explained that biometric information concerning facial characteristics was more sensitive than most other types of personal data.[46] Therefore, although there was no clear standard in the law regulating FRT at that time, the court held that HSP's use of this technology should be subject to more scrutiny.[47] Based on such understanding, the court ruled on 9 April 2021 that HSP was liable for using the plaintiff's facial data in the FRT systems without his consent.[48]

Some might believe that the political atmosphere was also favourable for Guo. While the Hangzhou Intermediate People's Court was hearing the case, the National People's Congress passed the Civil Code on 28 May 2020, with personal information protection as one of its salient points. China Central Television, the nation's largest state broadcaster, collaborated with China's Supreme People's Court and showcased this case as one of the ten benchmark cases in 2021.[49] Official publications by China's judiciary likewise prized the case as a sign of a progressive, more benevolent legal system.[50]

---

[42] Ibid.

[43] See Guobing Su Hangzhou Yesheng Dongwushijie Youxian Gongsi Fuwu Hetong Jiufen An (郭兵诉杭州野生动物世界有限公司服务合同纠纷案) [*Guo Bing v. Hangzhou Safari Park Co., Ltd.*], Hangzhou Fuyang District People's Court Case No. (2019) Zhe 0111 Minchu 6971, 20 November 2020.

[44] Ibid.

[45] Ibid.

[46] Guobing Su Hangzhou Yesheng Dongwushijie Youxian Gongsi Fuwu Hetong Jiufen An (郭兵诉杭州野生动物世界有限公司服务合同纠纷案) [*Guo Bing v. Hangzhou Safari Park Co., Ltd.*], Hangzhou Interm. People's Ct. of Zhejiang Province Case No. (2020) Zhe 01 Minzhong 10940, 9 April 2021.

[47] Ibid.

[48] Ibid.

[49] See, e.g., *China Daily*, 'Xin Shidai Tuidong Fazhi Jincheng 2021 Niandu Shida Anjian Jiexiao' (《"新时代推动法治进程2021年度十大案件"揭晓》) [Revealing ten cases of the year 2021 for the progress of the rule of law in the new era] (22 January 2022), https://cn.chinadaily.com.cn/a/202201/22/WS61ebd6caa3107be497a036f7.html.

[50] See, e.g., China Court, 'Renlian Shibie Jiufen Diyi An: Geren Xinxi Sifa Baohu De Dianfan' (《人脸识别第一案：个人信息司法保护的典范》) [The first court case involving facial recognition

Nevertheless, Guo himself was not satisfied with the judgment. He argued that the use of FRT by HSP was illegal per se,[51] but this viewpoint was not accepted by the court. Given the pervasive FRT in China, agreeing with Guo could be a step too far.

### 17.3.2 *The Beijing Metro Station*

In January 2022, Tsinghua law professor Dongyan Lao posted a long essay about China's social and legal problems on Weibo – the Chinese equivalent of Twitter.[52] One thing Lao lamented was her failed attempt to prevent the use of FRT in Beijing's subway stations.[53]

When the Beijing Subway Limited Company proposed to implement FRT in its 'real-name-based passenger' system, Lao was among the first against it.[54] In 2019, the Beijing's Rail Transit Control Centre, which is the administrative body responsible for underground transport in Beijing, announced the plan of enhancing subway station security by building an FRT-based railway passenger classification system.[55] The Centre explained that this system would not only protect public security of the Beijing subway, but also promote traffic efficiency.[56] The system was based on an AI-enabled facial image database, which could push security alerts automatically to personnel on site and drastically lessen their workloads.[57]

Shortly after the announcement, Lao openly expressed concerns regarding the over-intrusiveness of FRT in public venues and questioned the justification of this decision.[58] While China did not have any legislation regulating the FRT at that time, Lao argued that the rail transit agency had no authority to make such a

---

technology: A judicial epitome for personal information protection] (8 March 2022), www.chinacourt.org/article/detail/2022/03/id/6562816.shtml.

[51] See, e.g., Ye Yuan, 'A professor, a zoo, and the future of facial recognition in China' (26 April 2021), *Sixth Tone*, www.sixthtone.com/news/1007300/a-professor%2C-a-zoo%2C-and-the-future-of-facial-recognition-in-china.

[52] See David Cowhig, '2022: Chinese law prof's lament and encouragement' (29 January 2022), David Cowhig's Translation Blog, https://gaodawei.wordpress.com/2022/01/29/2022-chinese-law-profs-lament-and-encouragement/.

[53] Ibid.

[54] See Jeffrey Ding, 'ChinAI #77: A strong argument against facial recognition in the Beijing subway' (10 December 2019), *ChinAI Newsletter*, https://chinai.substack.com/p/chinai-77-a-strong-argument-against.

[55] Masha Borak, 'Beijing's subway system will use facial recognition to single out people for different security measures' (1 November 2019), *South China Morning Post*, www.scmp.com/abacus/tech/article/3035661/beijings-subway-system-will-use-facial-recognition-single-out-people.

[56] See Jeffrey Ding's translation of Lao's post at Ding, ChinAI #77.

[57] See *Beijing News*, 'Beijing Ditie Youwang Yingyong Renlian Shibie Jishu' (《北京地铁安检有望应用人脸识别技术》) [Beijing Metro security checks set to adopt facial recognition technology] (30 October 2019), http://epaper.bjnews.com.cn/html/2019-10/30/content_769638.htm?div=0.

[58] See Jeffrey Ding's blog: Ding, ChinAI #77

decision without conducting a public hearing.[59] In addition, Lao indicated that the system treated all passengers as potential criminals and therefore violated the presumption of innocence doctrine, which is fundamental to any modern criminal law system.[60] Shortly after this criticism, Lao's Weibo account was suspended and her posts were no longer available.[61]

To Lao's dismay, although the Centre postponed the plan of implementing FRT for nearly two years, it started to introduce the system in several stations in 2022.[62] The Centre compromised by adopting the FRT-based system on a voluntary basis. Passengers could get an express pass by completing real-name registration and uploading their facial data.[63] Beijing municipal government explained that the facial data was also linked to vaccination and testing results for the purpose of pandemic control. The Beijing municipal government announced in May 2022 that the system would be further linked to China's 'health code' – the mobile application used by Chinese people for mandatory checks on location data as well as COVID-19 testing reports.[64] Linking facial data to other types of sensitive personal information such as one's records of geo-location, could construe a form of highly aggregated data profiling. Information that does not seem to pose immediate harm might be less innocuous once a person's social relationships and patterns of behaviour are revealed through an extended period of data collection and aggregation. This aggregation problem can lead to highly intrusive portrayals of an individual's intimate life details, posing a unique threat to one's privacy. Lao's case reveals that the use of FRT for public security purposes can be easily justified by the authority and that challenging the government's use of FRT can face unsurmountable difficulties.

## 17.4 FRT IN THE SURVEILLANCE STATE

Although the Civil Code and PIPL have advanced personal data protection in China, Sections 17.2 and 17.3 have revealed that FRT used by the public sector has not been subject to much limitation. The government can always justify such use for the purpose of public security. This asymmetric regulatory model is rooted in China's unique political economy and regulatory philosophy.

---

[59] Ibid. for Ding's translation.
[60] Ibid.
[61] See, e.g., Stella Chen, 'Weibo chairman backs Chinese censor's crackdown and promises "ecologically sound" cyberspace' (25 September 2022), *South China Morning Post*, www.scmp.com/news/china/politics/article/3193605/weibo-chairman-backs-chinese-censors-crackdown-and-promises.
[62] See Cowhig's translation of Lao's essay: Cowhig, 'Chinese law prof's lament'.
[63] See *Southern Metropolis Daily*, 'Beijing Ditie Youjian Shualian Anjian, Yin Yinsi Xielu Danyou Zhuanjia: Yingxian Zhengqiu Yijian' (《北京地铁又见刷脸安检，引隐私泄露担忧 专家：应先征求意见》) [Beijing Metro resorts to facial recognition for security checks, causing concerns for data leaks. Experts: should consult the public's opinion] (29 December 2021), *Southern Metropolis Daily*, https://m.mp.oeeee.com/a/BAAFRD000020211229638893.html.
[64] See, e.g., Coco Feng, 'Coronavirus: Beijing, fighting Omicron, adds identity info to transport passes to speed up checks of Covid-19 status' (18 May 2022), *South China Morning Post*, www.scmp.com/tech/article/3178195/coronavirus-beijing-fighting-omicron-adds-identity-info-transport-passes-speed.

First, the asymmetric regulatory model has been hugely influenced by China's unique human rights values. The fundamentals of China's human rights are different from those of the Western world. In the Western world, human rights were designed to protect individuals from state power from the beginning.[65] However, China has viewed human rights as derived from the state, which reigns supreme over the individual.[66] Consequently, China's approach to human rights has been largely state-centric and emphasises individual responsibilities over individual rights.[67] Privacy is no exception. China's data protection philosophy is built on the view that data collection and analysis should be actively cultivated to boost state capacity to achieve a wide range of social governance objectives.[68] Although the law provides citizens with considerable protection for their data privacy, it also creates numerous opportunities for the government to infringe upon citizens' privacy. This understanding well explains why the public security interest, which is usually represented by the government, is always superior to personal data rights.

Second, Chinese law's tolerance of FRT is closely related to its real-name registration policy. While anonymity is an important instrument to promote citizens' free speech and to protect them against government retribution in many countries,[69] the Chinese government has strictly enforced a nationwide 'real-name registration' policy to maintain social and political stability by eliminating digital anonymity.[70] Under this policy, Chinese authorities have required users to register their real identities with internet and telecommunications service providers when using their services through various authentication mechanisms for easy traceability since the early 2000s.[71] The wide adoption of FRT has been a natural development to streamline the enforcement of the real-name registration policy because this technology has become the most efficient and effective identity verification technique.[72] Mobile users, for example, are required to register through facial scanning when buying new SIM cards.[73]

---

[65] Jyh-An Lee, 'Hacking into China's cybersecurity law' (2018) 53 *Wake Forest L. Rev.* 57–104, at 99–100.
[66] Ibid., 100.
[67] Ibid.
[68] William Chaskes, 'The three laws: The Chinese Communist Party throws down the data regulation gauntlet' (2022) 79 *Wash. & Lee L. Rev* 1169–1224, at 1182–1184.
[69] Christopher Slobogin, 'Public privacy: Camera surveillance of public places and the right to anonymity' (2002) 72 *Miss. L.J.* 213–315, at 240–243.
[70] See Lee and Liu, 'Real-name registration rules', pp. 11–15.
[71] Ibid.
[72] Elizabeth A. Rowe, 'Regulating facial recognition technology in the private sector' (2020) 24 *Stan. Tech. L. Rev.* 1–54, at 23–24.
[73] See Lily Kuo, 'China brings in mandatory facial recognition for mobile phone users' (2 December 2019), *The Guardian*, www.theguardian.com/world/2019/dec/02/china-brings-in-mandatory-facial-recognition-for-mobile-phone-users.

Third, China is an unparalleled surveillance state extensively using digital technologies to maintain its regime. Personal data, including facial data, is a key resource for the Chinese government to implement its ambitious national plans towards an algorithmically governed socialist state.[74] The collection and processing of facial data has become increasingly essential for the government to build an effective surveillance system and to carry out economic plans, such as the ambitious 'smart city' initiative.[75] According to a recent report analysing more than 100,000 government bidding documents from China, one FRT-based project in Fujian Province alone could produce more than 2.5 billion images to be stored by the police in the cloud at any given time.[76] Given the extensive integration of FRT in public infrastructures, it is unlikely that the Chinese judiciary and government would easily declare such use illegal or unjustified. Similarly, it will be too costly for the legislators to roll back FRT deployment prescribed by other branches of the authorities.[77]

## 17.5 CONCLUSION

With the enactment of the Civil Code and PIPL, China has substantially enhanced its personal data protection. According to these two laws and the Judicial Interpretation on FRT, facial data is defined as sensitive personal information, and the deployment of FRT is more restrictive. The case of *HSP* represents the country's determination to prevent the over-use of facial data in the private sector. However, China still faces serious challenges regarding FRT-related personal data protection under its asymmetric regulatory framework. While the use of FRT is increasingly regulated in the country, the regulatory restrictions can be invariably lifted for the reason of public security. Government agencies have invariably claimed this regulatory exemption for its massive FRT deployment. Moreover, the liability for the government's abuse or misuse of personal data is quite insignificant compared with that that for private parties. This asymmetric framework has resulted from China's unique human rights philosophy, the endeavour to enforce a real-name registration policy, and, more importantly, its determination to sustain a digital surveillance state.

---

[74] Ira S. Rubinstein, Gregory T. Nojeim, and Ronald D. Lee, 'Systematic government access to personal data: a comparative analysis' (2014) 4(2) *International Data Privacy Law* 96–119, at 98, https://doi.org/10.1093/idpl/ipu004; Kevin Werbach, 'Orwell that ends well? Social credit as regulation for the algorithmic age' 2022 (4) *U. Ill. L. Rev*, 1417–1475, at 1427–1431.

[75] Givens and Lam, 'Smarter cities or Bigger Brother?', 851–858.

[76] Isabelle Qian, Muyi Xiao, Paul Mozur, and Alexander Cardia, 'Four takeaways from a *Times* investigation into China's expanding surveillance state' (21 June 2022), *New York Times*, www.nytimes.com/2022/06/21/world/asia/china-surveillance-investigation.html.

[77] See Luo and Guo, 'Facial recognition in China', 178.